\documentclass[aip,cha,reprint]{revtex4-1}

\usepackage[pdftex]{graphicx}
\usepackage{epstopdf}
\usepackage{enumitem}
\usepackage{amssymb,amsfonts,amsmath}

\begin{document}

%\preprint{}

\title{A classification scheme for chimera states} %Title of paper



\author{Felix P. Kemeth}
%\thanks{}
\affiliation{Physik-Department, Nonequilibrium Chemical Physics, Technische Universit\"{a}t M\"{u}nchen,
  James-Franck-Str. 1, D-85748 Garching, Germany}
\affiliation{Institute for Advanced Study - Technische Universit\"{a}t M\"{u}nchen,
  Lichtenbergstr. 2a, D-85748 Garching, Germany}
\affiliation{The Department of Chemical and Biological
  Engineering - Princeton University, Princeton, NJ 08544, USA}

\author{Sindre W. Haugland}
\affiliation{Physik-Department, Nonequilibrium Chemical Physics, Technische Universit\"{a}t M\"{u}nchen,
  James-Franck-Str. 1, D-85748 Garching, Germany}
\affiliation{Institute for Advanced Study - Technische Universit\"{a}t M\"{u}nchen,
  Lichtenbergstr. 2a, D-85748 Garching, Germany}

\author{Lennart Schmidt}
\affiliation{Physik-Department, Nonequilibrium Chemical Physics, Technische Universit\"{a}t M\"{u}nchen,
  James-Franck-Str. 1, D-85748 Garching, Germany}

\author{Ioannis G. Kevrekidis}
\affiliation{Institute for Advanced Study - Technische Universit\"{a}t M\"{u}nchen,
  Lichtenbergstr. 2a, D-85748 Garching, Germany}
\affiliation{The Department of Chemical and Biological
  Engineering - Princeton University, Princeton, NJ 08544, USA}

\author{Katharina Krischer}
\email[]{krischer@tum.de}
\affiliation{Physik-Department, Nonequilibrium Chemical Physics, Technische Universit\"{a}t M\"{u}nchen,
  James-Franck-Str. 1, D-85748 Garching, Germany}


% \date{\today}

\begin{abstract}
We present a universal characterization scheme for chimera states
applicable to both numerical and experimental data sets. 
The scheme is based on two correlation measures that enable a meaningful
definition of chimera states as well as their classification into
three categories:
\textit{stationary}, \textit{turbulent} and \textit{breathing}. In addition, 
these categories can be further subdivided according to the
time-stationarity of these two measures.
We demonstrate that this approach both is consistent with previously recognized chimera states
and enables us to
classify states as chimeras which have not been
categorized as such before.
Furthermore, the scheme allows for a qualitative
and quantitative comparison
of experimental chimeras with chimeras obtained through
numerical simulations.
\end{abstract}


\maketitle %\maketitle must follow title, authors, abstract and \pacs

\begin{quotation}
The paper ``Coexistence of Coherence and Incoherence in Nonlocally
Coupled Phase Oscillators'' by Kuramoto and Battogtokh published in
2002\cite{kuramoto2002} marks the commencement of intense research
activities  on a counter-intuitive phenomenon that has come to be known as
a chimera state\cite{abrams2004}, i.e., the coexistence of coherent
and incoherent dynamics in a network of symmetrically coupled identical oscillators. 
For a
long time, the coexistence of coherence and incoherence had been
believed to be bound to heterogeneous networks of oscillators, in which
oscillators with a similar frequency might mutually synchronize,
while those with larger deviations of their frequencies from the mean
frequency keep on drifting incoherently. 
The discovery
that an array of identical oscillators, all coupled in an
identical way to their neighbors, can also be split into synchronized and
drifting groups was likewise surprising as fundamental.
The chimera state, being a novel type of dynamic state, can broaden
our understanding of transitions from synchrony to ``turbulence'' and vice
versa, and has possible realizations and applications in nature,
e.g. in neuroscience\cite{laing,Rattenborg2000} or
hydrodynamics\cite{Barkley2005,Duguet2013}. 
Since the pioneering works in the early years of this millennium,
chimera states have been observed in many different systems, ranging
from systems with non-local
coupling\cite{bordyugov2010,omelchenko2011,hagerstrom2012,omelchenk2013,sethia2013,zakharova2014}, 
via two-group
approximations\cite{abrams2008,tinsley2012} to global all-to-all
coupling\cite{schmidt2014,yeldesbay2014}. 
Due to
their robustness to noise, chimera states have also been observed
experimentally, e.g. in networks of coupled chemical
oscillators\cite{tinsley2012}, 
arrays of coupled spatial light modulators\cite{hagerstrom2012},
networks of mechanical oscillators\cite{martens} and
electrochemical systems\cite{schmidt2014}.
However, the various systems differ strongly in the visual attributes
of their dynamic behavior, asking for a systematic categorization.
In this paper, we propose a classification scheme based on linear
methods, which we believe fulfills the requirements of being universal
and simple in its application.
\end{quotation}
\section{Introduction}
Most early studies on chimera states dealt with non-locally coupled
phase oscillators, where coherence refers to phase- and frequency-locked
oscillators and incoherence to drifting oscillators, respectively
\cite{kuramoto2002}. 
Lately, more and more chimera patterns were discovered,
wherein coherence and incoherence is of a different nature.
One example is a coupled-map chimera, where the individual elements
consist of period-two
orbits. The coexistence pattern is composed of two synchronous
regions corresponding to the two realizations of the period-two orbit,
with a spatially incoherent interfacial region, where the spatial
arrangement of the two states appear in a random and thus incoherent
manner\cite{omelchenko2011}. Yet, each state remains a period-2 orbit in
time and is thus either synchronized or anti-synchronized to any of
the other elements, preserving temporal order. 
Another example is the so-called amplitude chimera,
where the incoherent group is
characterized by disorder in the amplitude of the oscillators while all
the oscillators
in the entire ensemble oscillate with the same
frequency\cite{zakharova2014}. 
Other coherence/incoherence coexistence
patterns differ from the classical chimera state by the variability
of coherent and incoherent regions,
which might both change their sizes and move in
space\cite{Battogtokh1997,schmidt2015}. 
Furthermore, the stability properties of these diverse
chimera states vary greatly. Many chimeras, among them the
original one in systems of nonlocally coupled phase oscillators, are
transient for a finite number of oscillators, but have a diverging
transient time in the
continuum limit $N \rightarrow \infty$\cite{wolfrum2011}. Others are stable already from small
ensemble sizes on \cite{ashwin2015,Schmidt2015_2}, and still others
have finite transient times
even in the continuum limit\cite{zakharova2014}.\par
These examples illustrate that the original definition of a chimera
state as ``a spatio-temporal pattern in which a system of identical
oscillators is split into coexisting regions of coherent and
incoherent oscillators''\cite{panaggio2015} 
does not cope with recent developments
but calls for a more distinct characterization and refinement.
There already exist two approaches
towards characterization schemes in minimal networks\cite{ashwin2015} and for
chimeras with non-local coupling\cite{gopal2014}, but they are both restricted to a
small class of systems.\par
In this paper we propose two measures for characterizing chimera states.
Although based on linear methods, these quantities provide what we believe to be a clear and simple definition of chimera states, and, furthermore, 
they allow for an easy distinction between chimera states with
different coherence properties and thus provide a useful classification scheme.
In addition, our approach is independent of the coupling scheme and the spatial
dimension of the system, and not restricted to phase oscillators, such
as the (local) Kuramoto order parameter\cite{kuramoto2002}.\par
The paper is structured as follows: In section II we introduce a spatial and
a temporal correlation measure applicable to arbitrary data sets and
define chimera states with the help of these measures. 
In section III, these criteria are applied to experimental and simulated data
of different chimera states, and in section IV a detailed
characterization scheme on the basis of the measures is discussed.
Details pertaining to the individual systems and to the numerical
methods are given in the Supplementary Information (SI)\cite{supp}.
\section{Correlation measures for spatial and temporal coherence}
\subsection{A measure for correlation in space}
\begin{figure}[htc]
  \centering
  \includegraphics[]{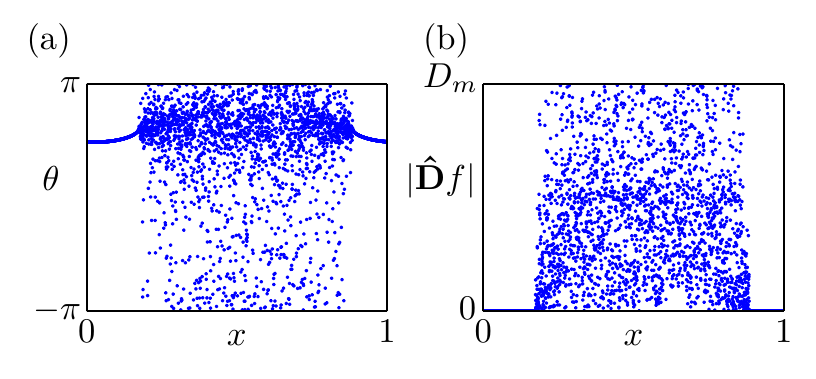}
  \caption{(a) Snapshot of the Kuramoto model\cite{kuramoto2002}, SI
    section I\cite{supp}, after the initial conditions decayed.
(b) Absolute value of the local curvature obtained by applying the discrete Laplace operator on the data set shown in (a).}
  \label{fig:1}
\end{figure}

For systems with a spatial extent, that is, systems with a local or
non-local coupling topology,
we employ the local curvature as a
measure for the spatial coherence.
Hereby the local curvature of the observable is quantified by the
second derivative in one-dimension, or, more generally, by the
Laplacian for any number of spatial dimensions.
Therefore, we calculate the local curvature at each point in space
by re-scaling and applying
the discrete Laplacian $\mathbf{D}$ on each snapshot
containing the spatial data $f$. For one snapshot at time $t$
with one
spatial dimension, this operation reads
\begin{equation}
\begin{split}
  \label{eq:1}
  \mathbf{\hat{D}}f & = \Delta x^2 \mathbf{D}f \\
  & = f \left(x+ \Delta x,t\right)-2f \left(x,t\right)+f \left(x- \Delta x,t \right),
\end{split}
\end{equation}
where each data point in $f$ can be either real, complex or of any higher dimension.
In order to clarify this concept, 
consider the chimera state observed by Kuramoto and Battogtokh in a
ring of non-locally
coupled phase oscillators\cite{kuramoto2002}. 
One realization of the chimera state is depicted in figure
\ref{fig:1}a.
Through the application of the discrete Laplace operator,
this snapshot is mapped onto a new function as shown in figure
\ref{fig:1}b, with $D_{m}$ indicating the maximal value of
$|\mathbf{\hat{D}}|$. 
Note that for phase oscillator systems, we apply this operator on the data in
the complex plane, that is the phase oscillators are located on a ring
with  constant amplitude $A$. Then, $D_{m}$ corresponds to the curvature at 
an oscillator whose two neighbors are shifted 
$180^{\circ}$ in phase, i.e., whose neighboring oscillators are located on opposite
positions on the circle. With the re-scaling obtained by multiplying
the Laplacian with $\Delta x^2$ in Eq.~\eqref{eq:1}, $D_{m}$ converges to
$4A$ in the continuum limit. In the synchronous regime $\lim_{N \rightarrow \infty}
|\mathbf{\hat{D}}| = 0$. This means
that the synchronous regime is projected onto the x-axis through this
transformation, while in the incoherent regime $|\mathbf{\hat{D}}|$ is
finite and exhibits pronounced fluctuations. 
Consequently, when we consider the normalized probability density function
$g$ of $|\mathbf{\hat{D}}|$,
$g(|\mathbf{\hat{D}}|=0)$
measures the relative size of spatially coherent regions in each
temporal realization. For a fully synchronized system
$g(|\mathbf{\hat{D}}|=0)=1$, while a totally incoherent system gives a
value $g(|\mathbf{\hat{D}}|=0)=0$. A value between $0$ and $1$ of
$g(|\mathbf{\hat{D}}|=0)$ indicates coexistence of synchrony and incoherence.\par
Given this discussion, two important aspects have to be considered.
First, the definition of spatial coherence and incoherence is not
absolute,
but has to be compared to the maximal curvature in each system.
Thus, we argue that the characterization of coherence and incoherence is relative and
depends on the individual system.
Second, even in the coherent region, there might be some minor change
in state (cf. figure \ref{fig:1}a above) leading to a non-zero
curvature.
Hence, we are convinced that in order to characterize something as
coherent or incoherent, 
a threshold value is inevitable, although, as will be shown later, the exact position of the threshold does not
change the qualitative outcome.\par
Considering the two arguments above, we propose that for spatially
extended systems, 
a point for which the absolute local curvature is less than one percent of the
% a point with an absolute curvature less than one percent of the
maximum curvature present in the system
should be characterized as coherent, and as incoherent otherwise.\par
\begin{figure}[htc]
  \centering
  \includegraphics[]{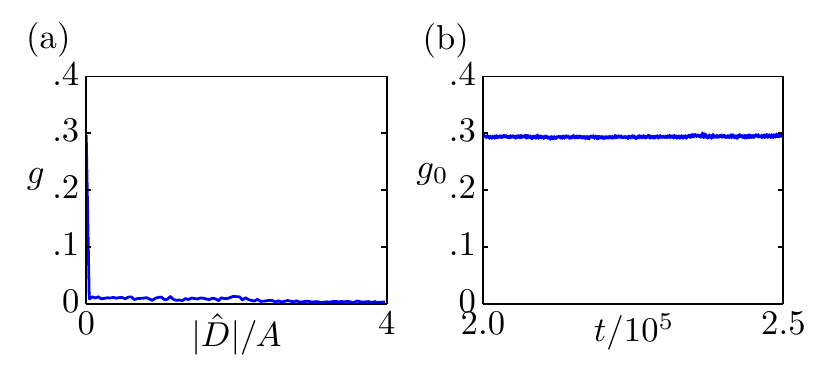}
  \caption{(a) Probability distribution function $g$ of the discrete
    Laplace operator applied on the snapshot of figure \ref{fig:1}a.
(b) Temporal evolution of $g_0(t)$ for a longer time series of the
Kuramoto model.}
  \label{fig:2}
\end{figure}

With the
threshold $\delta = 0.01D_{m}$
our first correlation measure
\begin{equation}
  \label{eq:5}
  g_0(t) := {\textstyle \int}_0^{\delta} g(t,|\hat{D}|) d|\hat{D}|
\end{equation}
can be used to
% describe the amount of spatially coherent
describe the spatial extent occupied by coherent
oscillators, even for systems beyond coupled
phase oscillators. An example of $g$ for the Kuramoto model is shown
in figure \ref{fig:2}a. Note that, in general, $g$ is time dependent.
Figure \ref{fig:2}b shows $g_{0}(t)$ as a function of time. The
value of $g_{0}(t)$ of about 0.3 confirms the interpretation of the
state as a chimera state, 
% while its visual time-independence
% strongly suggests that the
while its time-independence reveals that the
degree of coherence is stationary.\par
For systems without a spatial dimension, i.e., systems with solely global
coupling, curvature is not defined. Nevertheless, we argue that the
pairwise Euclidean distances between the values of all oscillators, $f_i$,
\begin{equation}
  \label{eq:2}
  \mathbf{\tilde{D}} = \{\tilde{D}_{ij}\} = \| f_i - f_j \| \, , \, i \neq j,
\end{equation}
are a good measure for synchrony/asynchrony.
Again, from the normalized
probability density function $g$ of $\mathbf{\tilde{D}}$, a variable
\begin{equation}
  \label{eq:7}
  \tilde{g}_0(t) :=\sqrt{{\textstyle \int}_0^{\delta} g(t,|\tilde{D}|) d|\tilde{D}|},
\end{equation}
can be obtained that is a measure for the relative amount of
correlated oscillators.
Here, the square root arises due to the fact that by taking all
pairwise distances, the probability of oscillators
$i$ and $j$ both being in the synchronous cluster equals
$(N_0/N)^2$, with $N_0$ being the number of the synchronous
oscillators. Since both measures, $g_0(t)$ and $\tilde{g}_0(t)$, describe
the same property, that is, the degree of spatial synchronization of the system, we only use $g_0(t)$
as notation
in the following.\par
\begin{figure}[htc]
  \centering
  \includegraphics[]{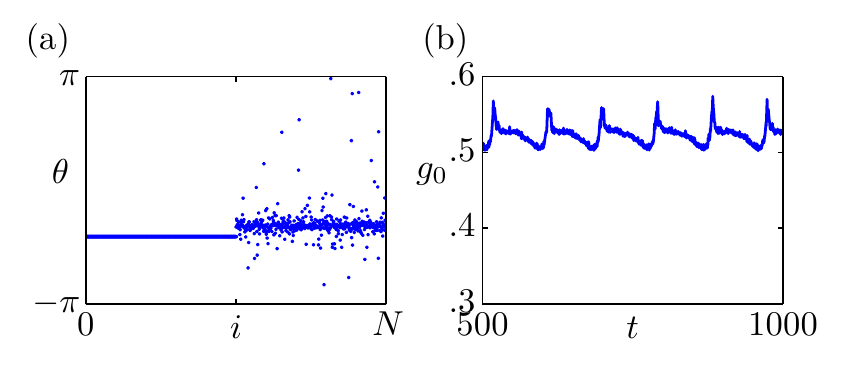}
  \caption{(a) Snapshot of a realization of the chimera state
    observed in the two-group model\cite{abrams2008}, SI section II\cite{supp}.
(b) Temporal evolution of $g_0(t)$.}
  \label{fig:3}
\end{figure}

As an illustration, consider the two groups of globally coupled
phase oscillators investigated by Abrams et al\cite{abrams2008}. An exemplary snapshot
is depicted in figure \ref{fig:3}a, where oscillators $1,\dots,N/2$
belong to group 1 and oscillators $N/2+1,\dots,N$ constitute
group 2. 
Clearly, group 1 is synchronous while the oscillators in
group 2 behave incoherently.
In the parameter region considered, a breathing of the chimera as
expressed through
an oscillation of the variance of the incoherent cluster was reported\cite{abrams2008}.
The temporal evolution of $g_0(t)$ is shown in figure \ref{fig:3}b.
It can be observed that $g_0(t)$, i.e., the relative amount of partially
synchronized oscillators,
evolves periodically and 'breathes' over time. Therefore, the temporal
evolution of $g_0(t)$ allows for the discrimination between chimeras
with constant and oscillating partial synchronization.
We term these \textit{stationary} and \textit{breathing} chimeras, respectively.
The latter term has been adapted from the literature,
since the Kuramoto order parameter $r$ exhibits the qualitatively same
temporal behavior as $g_0$\cite{abrams2008}.
Note that the two approaches above are independent of the spatial
dimension and the 
number of variables of the different systems. 
This makes $g_0(t)$ a versatile tool for the classification of multifaceted
data sets such as those obtained from chimera states.

\subsection{A measure for correlation in time}
\begin{figure}[htc]
  \centering
  \includegraphics[]{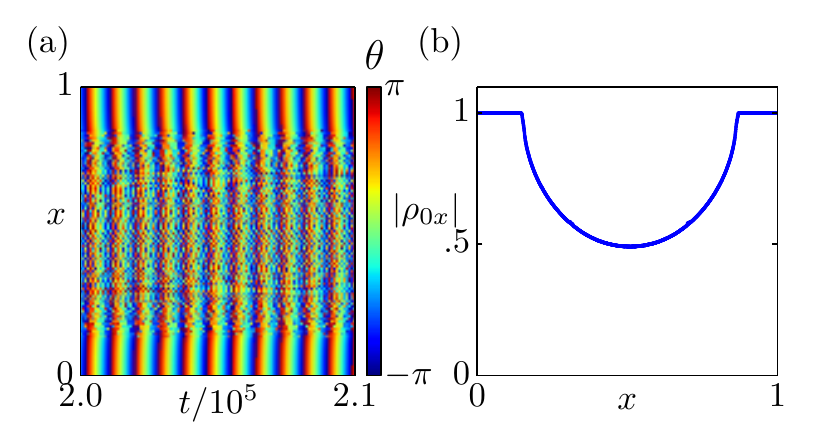}
  \caption{(a) Temporal evolution of the phase $\theta$ in the Kuramoto model\cite{kuramoto2002}, SI section I\cite{supp}. (b) Pairwise correlation coefficients $\rho_{0x}$ between
    the oscillator at $x=0$ and the remaining oscillators.}
  \label{fig:4}
\end{figure}

In addition to the measure for the spatial correlation discussed in the previous
section, the temporal correlation of the individual
oscillators provides valuable information for a distinction between
different chimera dynamics as well. 
Suppose
$X_i$ and $X_j$ are the real or complex time series of two individual
oscillators with $\mu_i,\mu_j$ and $\sigma_i,\sigma_j$ their
respective means and standard deviations.
Then, consider the pairwise correlation coefficients
\begin{equation}
  \label{eq:3}
  \rho_{ij} = \frac{\langle(X_i-\mu_i)^{\large{*}} \, (X_j-\mu_j) \rangle}{\sigma_{i}\sigma_{j}}
\end{equation}
with $\langle \cdot \rangle$ indicating the temporal mean and
$\large{*}$ complex conjugation. Note that
$\rho_{ij} = 1$ for linearly correlated time series, $\rho_{ij} = -1$
for linearly anti-correlated time series and $|\rho_{ij}| = 1, \angle
\rho_{ij} =
\alpha$ for complex time series with a constant phase shift of
$\alpha$. 
That means, the normalized distribution function $h$ of
\begin{equation}
  \label{eq:4}
  \mathbf{\hat{R}} = \{|\rho_{ij}|\} \, , \, i \neq j
\end{equation}
is a measure for the correlation in time.
For static chimera states, where the coherent cluster is localized at
the same position over time, $h(|\rho_{ij}| \approx 1)$ is non-zero.
In practice, we consider two oscillators as correlated if
$|\rho_{ij}| > 0.99=\gamma$. As an example, consider the
Kuramoto model mentioned above. Again, we map the system
onto the complex plane with arbitrary constant amplitude $A$ for all
oscillators. Then, for the chimera state depicted in figure
\ref{fig:4}a, we calculate the correlation matrix $\mathbf{\hat{R}}$
and its probability distribution function $h$. The first row of
$\mathbf{\hat{R}}$,
$\{\rho_{0x}\}$, is shown in figure \ref{fig:4}b. Note that
this approach maps the temporally coherent part onto 1, cf. figure
\ref{fig:4}b.
The distribution function $h$ is depicted in figure
\begin{figure}[htc]
  \centering
  \includegraphics[]{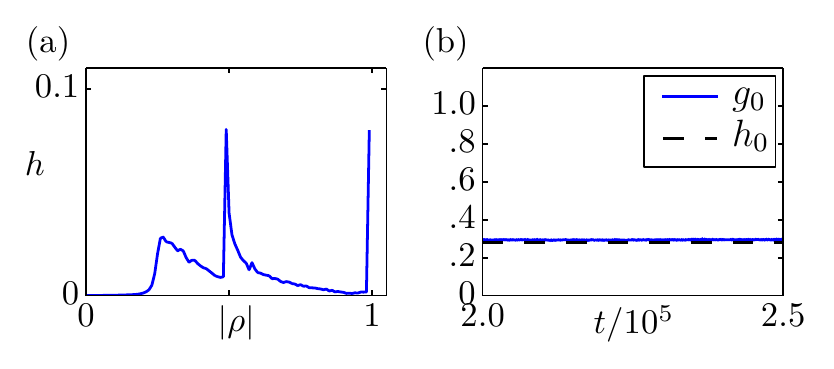}
  \caption{(a) Distribution function $h$. 
(b) Temporal evolution of $g_0(t)$ and the value of $h_0$, obtained from the same time interval. Note that
$h_0$ is not a function of time and is shown here only for comparison
with $g_0(t)$.}
  \label{fig:4b}
\end{figure}
\ref{fig:4b}a. It exhibits a distinct peak at $|\rho|=1$, indicating
that the chimera state is static, i.e., that the majority of oscillators
does not change its ``group affiliation''. We suggest to term this kind of
chimera state a
\textit{static} chimera.
The peak at
$|\rho|\approx 0.5$ arises due to the partial linear correlation
between oscillators at $x \approx 0.5$ and synchronous oscillators,
cf. figure \ref{fig:4}b.
The percentage of the time-correlated oscillators can now be
quantified with  
\begin{equation}
  \label{eq:6}
  h_0 := \sqrt{{\textstyle \int}_{\gamma}^1 h(|\rho|) d|\rho|}, 
\end{equation}
%%%%%%%%%%%%%%%%%%%%%%%%%%%%%%%%%%%%%%%%%%%%%%%%%%%%%%%%%
% e.g. $h_0 \approx \sqrt{0.08} \approx 0.28$ for the Kuramoto model,
% coinciding in this case, i.e., the case of a static chimera, with $g_0(t)$, see figure \ref{fig:4b}b.\par
e.g. $h_0 \approx \sqrt{0.08} \approx 0.28$ for the Kuramoto model,
see figure \ref{fig:4b}b.\par
%%%%%%%%%%%%%%%%%%%%%%%%%%%%%%%%%%%%%%%%%%%%%%%%%%%%%%%%
Note that $h_0$ does not always reflect the size of the
% It is not always the case that $h_0$ reflects the size of the
synchronized cluster. This is especially the case when coherent and
incoherent regimes are non-static and perform spatial movements
over time. Then, $h_0$ is much smaller than $g_0(t)$ and may vanish
for large enough time windows.
%%%%%%%%%%%%%%%%%%%%%%%%%%%%%%%%%%%%%%%%%%%%%%%%%%%%%%%%
$h_0$ coincides with $g_0$, cf. figure \ref{fig:4b}b, if and only if the chimera is static
and no spatial coherence is present in the incoherent cluster.

\section{Examples of chimera states and their characterization}
\begin{figure}[htc]
  \centering
  \includegraphics[]{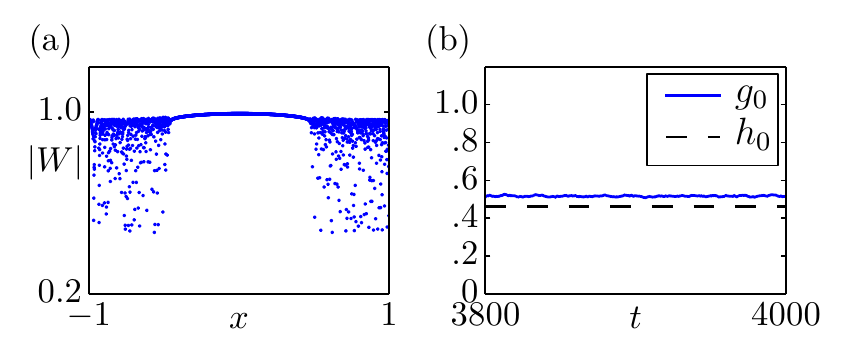}
  \caption{(a) Snapshot of the amplitude of the amplitude-mediated
    chimera\cite{sethia2013}, SI section III\cite{supp}. (b) $g_0(t)$ and $h_0$ of the amplitude-mediated chimera state.}
  \label{fig:5}
\end{figure}

As shown in the previous section, $g_0(t)$ of the Kuramoto model remains
constant in time and, in addition, coincides with $h_0$. This
indicates the constant phase relation between the coherent and incoherent
part and their spatial stationarity in time. The same
qualitative behavior can be observed in many different non-locally coupled
dynamical systems, such as
in non-locally coupled Stuart-Landau
oscillators investigated by Bordyugov et al.\cite{bordyugov2010} and in chimera states
observed by Sethia and Sen in a non-locally coupled version of the
complex Ginzburg-Landau equation (CGLE)\cite{sethia2013}. A snapshot and the observables
$g_0(t)$ and $h_0$ of the latter are depicted in figure \ref{fig:5}a
and b, respectively.
% If $g_0(t)$ is stationary and approximately coincides with $h_0$, independent of the
If $h_0$ is larger than 0, independent of the
size of the regarded time frame, then one can conclude that the
chimera state is stationary in the sense that the incoherent and
synchronous patches do not move. 
According to our definition above, this chimera state is
a \textit{static} chimera. Moreover, the finite values of $g_0(t)$ and $h_0$
indicate that the desynchronized dynamics are both \textit{spatially} and
\textit{temporally} incoherent.\par
\begin{figure}[htc]
  \centering
  \includegraphics[]{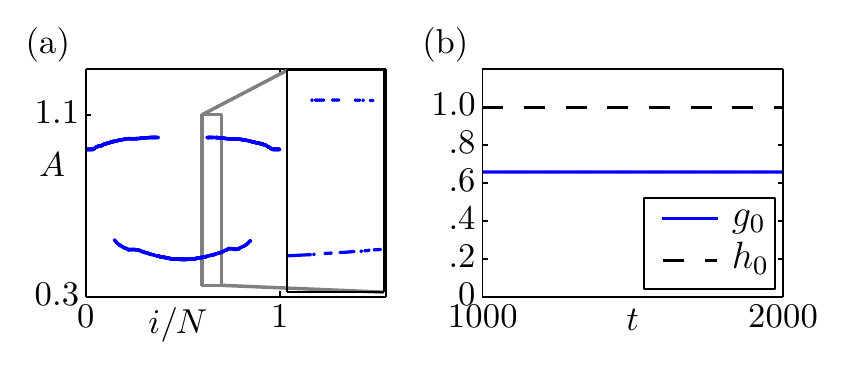}
  \caption{(a) Snapshot of the
    chimera state observed by Omelchenko et al.\cite{omelchenko2011}, SI section IV\cite{supp}.
    In the right part a magnification of the dynamics in the indicated
    rectangle is shown.
    (b) $g_0(t)$ and $h_0$ of
    the chimera state in (a).}
  \label{fig:6}
\end{figure}

An example of a static chimera state not exhibiting
temporal incoherence was examined by Omelchenko et al. in a system of
non-locally coupled maps with a period-2 orbit\cite{omelchenko2011}, and subsequently experimentally realized in Ref.~\cite{hagerstrom2012}. 
As depicted in figure \ref{fig:6}a, the
individual realizations are located on two stable branches. 
As evident from figure \ref{fig:6}b, for these chimeras $g_0(t)$ is
constant and smaller than $1$, while $h_0$ equals $1$. 
The value of $g_0(t)$ between 0 and 1 affirms that we are dealing with
a chimera state, while the fact that $h_0=1$ attests to the absence of
any temporal incoherence.\par
As already mentioned in the previous section, the temporal evolution
of $g_0(t)$ can be used to identify different dynamic behaviors of chimera states.
Apart from being constant, $g_0(t)$ can oscillate in time for a
\textit{breathing} chimera state, as
already shown in figure \ref{fig:3}b for the two-groups approximation\cite{abrams2008}. Another
example is the so-called type II chimera, which was reported in the CGLE
with nonlinear global coupling\cite{schmidt2014}.
\begin{figure}[htc]
  \centering
  \includegraphics[]{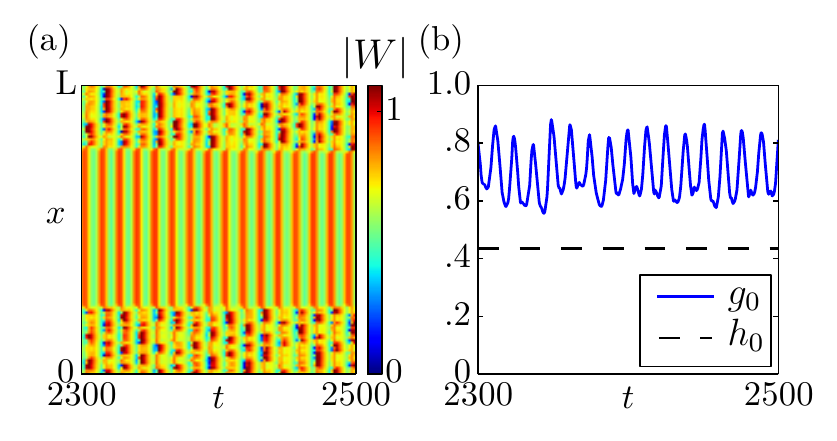}
  \caption{(a) Temporal evolution of the modulus of a one-dimensional simulation of the type II
    chimera state observed in the modified CGLE\cite{schmidt2014}, SI section VII\cite{supp}, with
    $L=1000$. (b) $g_0(t)$ and $h_0$ calculated from the data shown in (a).}
  \label{fig:7}
\end{figure}
The temporal evolution of the absolute value of the complex amplitude
and the observables $g_0(t)$ and $h_0$ are depicted
in figure \ref{fig:7}a and b, respectively. In figure \ref{fig:7}b, the oscillatory
behavior of $g_0(t)$ is evident, indicating
partial synchronization also in the incoherent regime. 
Note that within the incoherent cluster, there are
always homogeneous patches, leading to the offset between $g_0(t)$ and
$h_0$.\par
\begin{figure}[htc]
  \centering
  \includegraphics[]{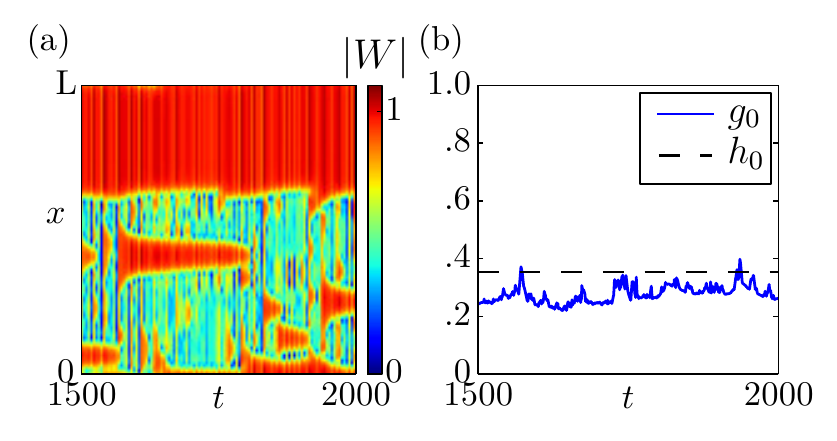}
  \caption{(a) Temporal evolution of the modulus of the complex amplitude of the type I
    chimera state observed in the CGLE with linear global
    coupling\cite{schmidt2015}, SI section VI\cite{supp}, with $L=200$. (b) $g_0(t)$ and $h_0$ calculated from the data shown in (a).}
  \label{fig:8}
\end{figure}

Besides oscillating in time, the observable $g_0(t)$ can also vary irregularly.
Such a behavior can be observed in the so-called type I chimera in the CGLE
with linear global coupling\cite{schmidt2015}. A representative evolution
of the modulus of the complex amplitude $W$ and the corresponding measures $g_0(t)$ and $h_0$ are depicted in
figure \ref{fig:8}a and b, respectively. Note that $h_0$ is significantly larger than 0, indicating
that the chimera state is static. The irregularity in $g_0(t)$ arises
from spatio-temporal intermittency, which appears spontaneously in the
turbulent regime, leading to 
the emergence of patches of oscillators that are synchronous with the coherent
region and shrink and disappear with time.
% a variation of the size of the homogeneous cluster.
Non-stationary chimera states with irregular phase boundaries
were also reported by Bordyugov et al.\cite{bordyugov2010}, who
named this state a \textit{turbulent} chimera. We adapt this expression for
general chimera states with irregular variation of the partial
synchronization, $g_0(t)$.\par
Dynamics resembling the type I chimera in some aspects is the
spatio-temporal intermittency as observed in
the CGLE\cite{shraiman1992}.
\begin{figure}[htc]
  \centering
  \includegraphics[]{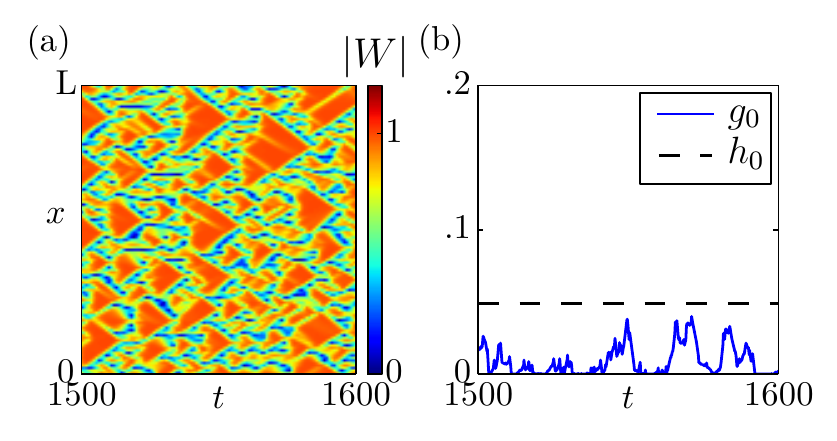}
  \caption{(a) Temporal evolution of 
    the modulus of the complex amplitude of the one-dimensional CGLE
    showing spatio-temporal intermittency\cite{shraiman1992}, SI section V\cite{supp}.
    (b) $g_0(t)$ and $h_0$ calculated from the data shown in (a).}
  \label{fig:9}
\end{figure}
A realization of the spatio-temporal intermittency in the
one-dimensional CGLE is shown
in figure \ref{fig:9}a. In figure \ref{fig:9}b, the irregular evolution of
$g_0(t)$ is apparent. However, 
in contrast to the type I chimera discussed above, $g_0(t)$ 
drops to zero at different points in time. This
means that the coherent part, and with it the coexistence between
synchrony and incoherence, vanishes completely from time to
time. 
% Therefore, spatio-temporal intermittency does not represent a chimera
% state in our view.
Therefore, spatio-temporal intermittency should not be considered to represent a chimera state.
$h_0$ is also small ($<0.05$), and results from the correlation
of neighboring points due to diffusion.\par
\begin{figure}[htc]
  \centering
  \includegraphics[]{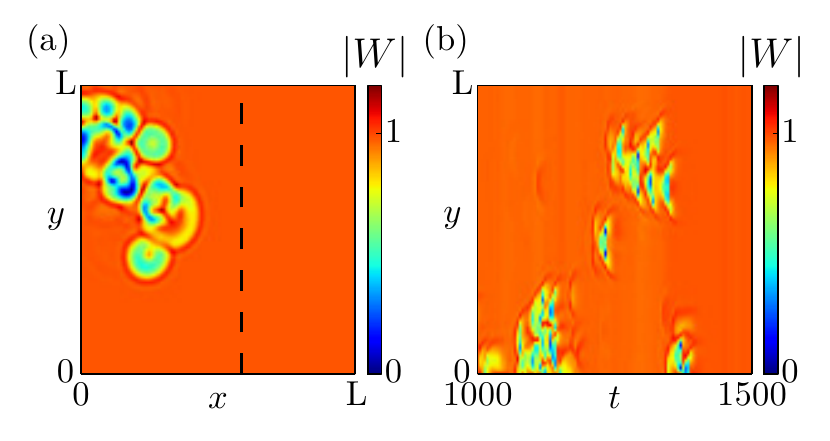}
  \caption{(a) Snapshot of the amplitude of the localized turbulence\cite{Battogtokh1997}
    at $t=1500$ with $L=200$, SI section VI\cite{supp}. (b) Temporal evolution of a one-dimensional cut
    at the $x$-value indicated by the dashed line in (a).}
  \label{fig:9b}
\end{figure}
\begin{figure}[htc]
  \centering
  \includegraphics[]{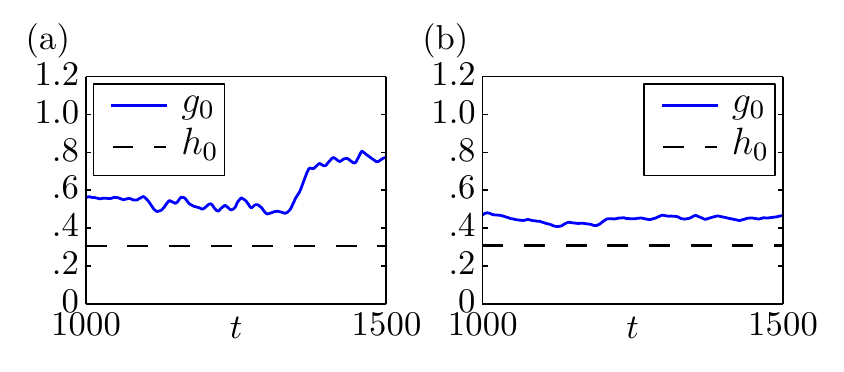}
  \caption{(a) $g_0(t)$ and $h_0$ of the localized turbulence with
    $L=200$. (b) $g_0(t)$ and $h_0$ of the localized turbulence with
    domain length $L=2000$.}
  \label{fig:9c}
\end{figure}

Dynamics with reversed roles, that is turbulent patches appearing in
an otherwise homogeneous regime, are found in the CGLE with linear\cite{Battogtokh1997} and
non-linear global coupling\cite{schmidt2015} and is called localized turbulence. An example
is shown in figure \ref{fig:9b}, with a snapshot of the modulus of a
two-dimensional simulation in (a)
and
the temporal evolution of a one-dimensional cut in (b).
The corresponding correlation measures $g_0(t)$ and $h_0$,
calculated from the two-dimensional spatio-temporal data with system
size $L=200$ are depicted in figure \ref{fig:9c}a.
The fluctuating value of $g_0(t)$ suggests that the degree of
coherence changes with time. A strong increase of the synchronous part
occurs at $t \approx 1350$, indicating a strong non-stationarity.
However, calculations with larger system sizes suggest that the
variations vanish in the thermodynamic limit $N \rightarrow \infty$.
An illustration is depicted in figure \ref{fig:9c}b, where $g_0(t)$
was calculated from two-dimensional simulations of systems with
$L=2000$.\par
A characteristic feature of localized turbulence, as compared to all chimera
states discussed above, is that the turbulent islands are composed of
several incoherent ``bubble-like'' structures, which move erratically in
the spatial domain. Bubbles disappear or pop up through division of
existing bubbles. Due to this steady motion of the turbulent islands,
the fraction of the coherent time series, as measured by $h_0$ is small, and vanishes if the time window is chosen large enough.
The same holds for the alternating chimeras observed by Haugland et
al.\cite{haugland2015}, where the turbulent part alternates with the homogeneous
regime in time (not shown).

\subsection{Transient chimeras}
\begin{figure}[htc]
  \centering
  \includegraphics[]{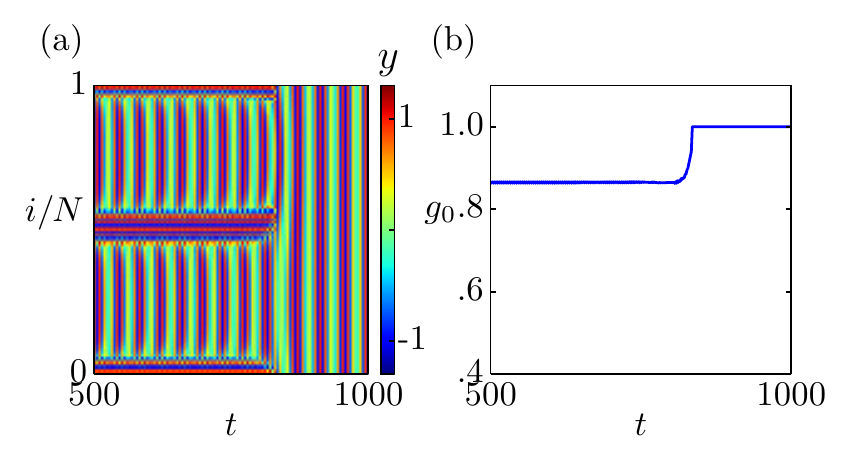}
  \caption{(a) Temporal evolution of the imaginary part, $y$, of the
    so-called amplitude chimera observed by Zakharova et al\cite{zakharova2014}, SI section VIII\cite{supp}. (b)
    $g_0(t)$ of the chimera shown in (a).}
  \label{fig:10}
\end{figure}

So far, we did not consider the long-term stability of the chimera
states yet. However, especially in the context of chimera states,
defining a stability concept is an important issue.
While various chimera states, as the type I and type II chimeras mentioned
above, are the only attractors for a specific parameter region,
and as such are stable, many other chimera states including those of the
Kuramoto model,
are long-term transients with infinite transient time in the continuum limit $N\rightarrow \infty$\cite{wolfrum2011}.
% Still, it could be shown that
% they become stable in 
% the continuum limit $N\rightarrow \infty$\cite{wolfrum2011}.
Then, there exist
states encompassing coexistence of coherence and incoherence that collapse
to the homogeneous state after a finite time even for $N\rightarrow \infty$. An
example thereof is the so-called amplitude
chimera\cite{zakharova2014}.
The space-time realization of such a state is depicted in figure
\ref{fig:10}a, figure \ref{fig:10}b showing the evolution of $g_0(t)$.
Amplitude chimeras resemble the chimeras found in coupled period-2
maps (cf. figure \ref{fig:6}) insofar as they are composed of two
coherent domains with anti-phase behavior that are separated by a
spatially incoherent interfacial region. In the latter region, the
absolute values of the amplitudes vary erratically in space but each 
oscillator is strictly periodic with a frequency equal to the
frequency of the synchronous regions. The spatial incoherence renders
$g_0(t)$ smaller than 1. However, as investigated in detail by Loos et
al.\cite{zakharova2016} and also evident from figure \ref{fig:10}, the
chimera-like dynamics are not stable.
A transition to full synchronization can be
observed, i.e. $g_0(t)=1$ after a finite time interval. In this case, the
lifetime of the chimera state strongly depends on the choice of the
initial conditions
and asymptotically approaches a constant value in the continuum limit
\cite{zakharova2016}.\par
We consider it meaningful to discriminate between transient chimeras and
chimera states which are attractive in the continuum limit. Therefore we
suggest to introduce a separate class \textit{transient} chimeras for
states with $0<g_0(t)<1 \forall t<t_0$ and $g_0(t) = 1 \vee g_0(t)=0$
at some transient time $t_0$.\par
\begin{figure}[htc]
  \centering
  \includegraphics[]{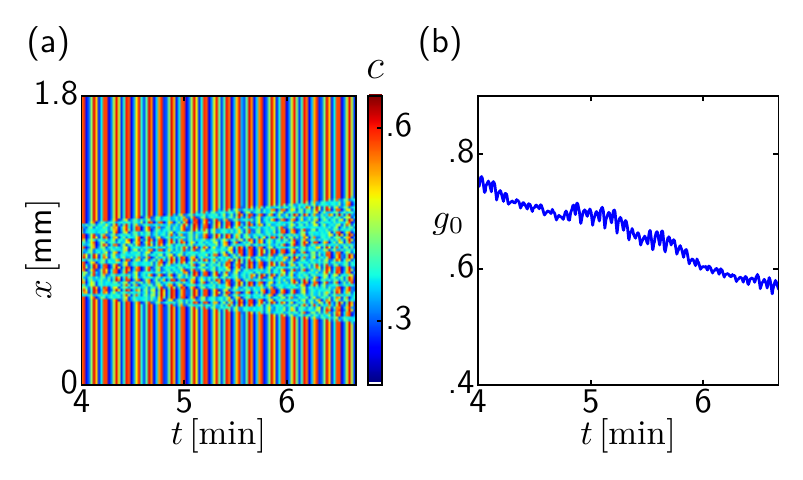}
  \caption{(a) Temporal evolution of the CO-coverage $c$, SI section IX\cite{supp}. (b) $g_0(t)$
    of the dynamics shown in (a).}
  \label{fig:11}
\end{figure}

Another remarkable case 
that created controversy as to its characterization as a chimera
% that was put up for debate whether the
% dynamics was chimera-like or not 
was reported by Falcke and Engel in a globally coupled
version of the CO-oxidation
model\cite{Falcke1993,Falcke1994,Falcke1995}. There, turbulent patches
appeared
in an otherwise homogeneously oscillating background, similar to the
localized turbulence discussed above. But, in contrast to the behavior
in the localized turbulence, no turbulent bubbles ever disappear.
A one-dimensional simulation
is depicted in \ref{fig:11}a, with the corresponding measure
$g_0(t)$ plotted in \ref{fig:11}b.
There, the incoherent region expands 
into the
synchronously oscillating domains
with an approximately constant velocity that 
is strongly dependent on the diffusion
coefficient $D$.
This non-stationarity manifests itself in the overall systematically declining
behavior of $g_0(t)$.
In such a case a longer simulation time is
necessary in order to verify that $g_0(t)$ vanishes after a finite time
interval, which was confirmed for the present case. Since it mediates
a transition from an unstable to a stable state, it fulfills the above
defined criteria for a \textit{transient} chimera state. 
We thus classify it accordingly.
% We suggest therefore to
% classify this behavior accordingly.
\begin{figure}[htc]
  \centering
  \includegraphics[]{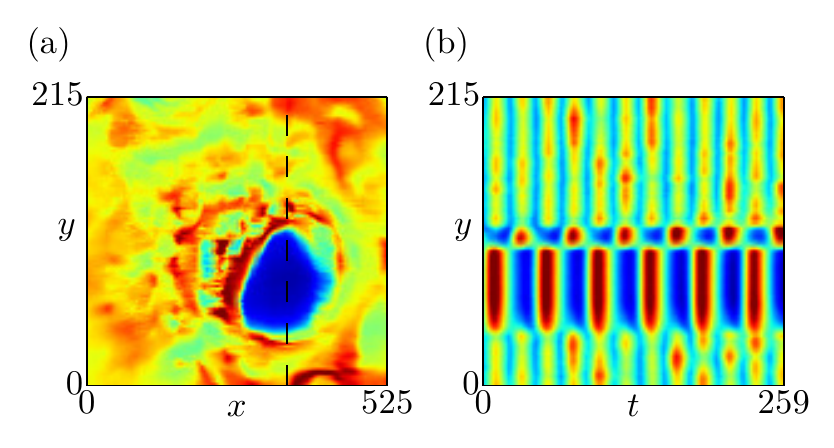}
  \caption{(a) Snapshot of the SiO$_2$ thickness on a Si-electrode
    in pseudo-colors. (b) Temporal evolution of a one-dimensional cut
    as shown in (a).}
  \label{fig:12}
\end{figure}

\subsection{Experimental observation of chimeras}
Chimeras have also been observed in experimental
setups\cite{hagerstrom2012,tinsley2012,schmidt2014}.
In this section, we apply our approach to experimental data as
described by Sch\"onleber et al\cite{Konrad2014}. 
In this system, the thickness of a SiO$_2$ layer on a Si-electrode
oscillates due to simultaneous electrochemical oxidation and etching.
% oscillates during electrochemical oxidation of
%the Si wafer.
Changes of the SiO$_2$ thickness
are measured via ellipsometric imaging.
A snapshot of a measurement is depicted in figure
\ref{fig:12}a,
with the color indicating the thickness of the oxide layer. The experimental
data was
processed using a moving average over the last $10$ time frames.
\begin{figure}[htc]
  \centering
  \includegraphics[]{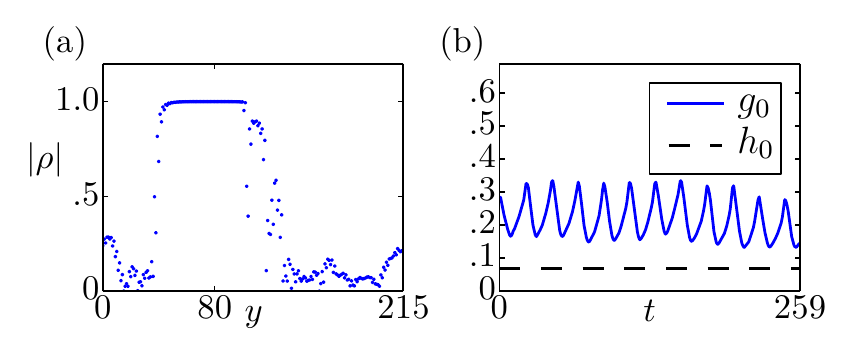}
  \caption{(a) Correlation coefficients for $y=80$ for the
    one-dimensional cut shown in \ref{fig:12}b. (b) $g_0(t)$ and $h_0$
    for the whole data set.}
  \label{fig:13}
\end{figure}
The temporal evolution of a one-dimensional cut is shown in figure
\ref{fig:12}b,
where the homogeneous oscillation of a small
region in an otherwise inhomogeneously oscillating background can be
observed. 
Figure \ref{fig:13}a shows
the pairwise correlation coefficients of the cross-section with a
point inside the coherent
cluster (here $y=80$): a strong linear correlation
within this cluster and the diminishing correlation with the remaining
oscillators is evident.
\begin{figure*}[htc]
\caption{Characterization of chimera states by means of $g_0(t)$ and
  $h_0$. The different examples of chimera states discussed in this paper are
  given in italics. In order to distinguish between no chimera and
  transient chimera, the transient time $t_0$ has to be much larger
  than the characteristic time of the uncoupled dynamics.}
\label{fig:tree}
\centering
\includegraphics[keepaspectratio=false,clip=true,trim=120px 500px 120px 120px,width=0.9\textwidth]{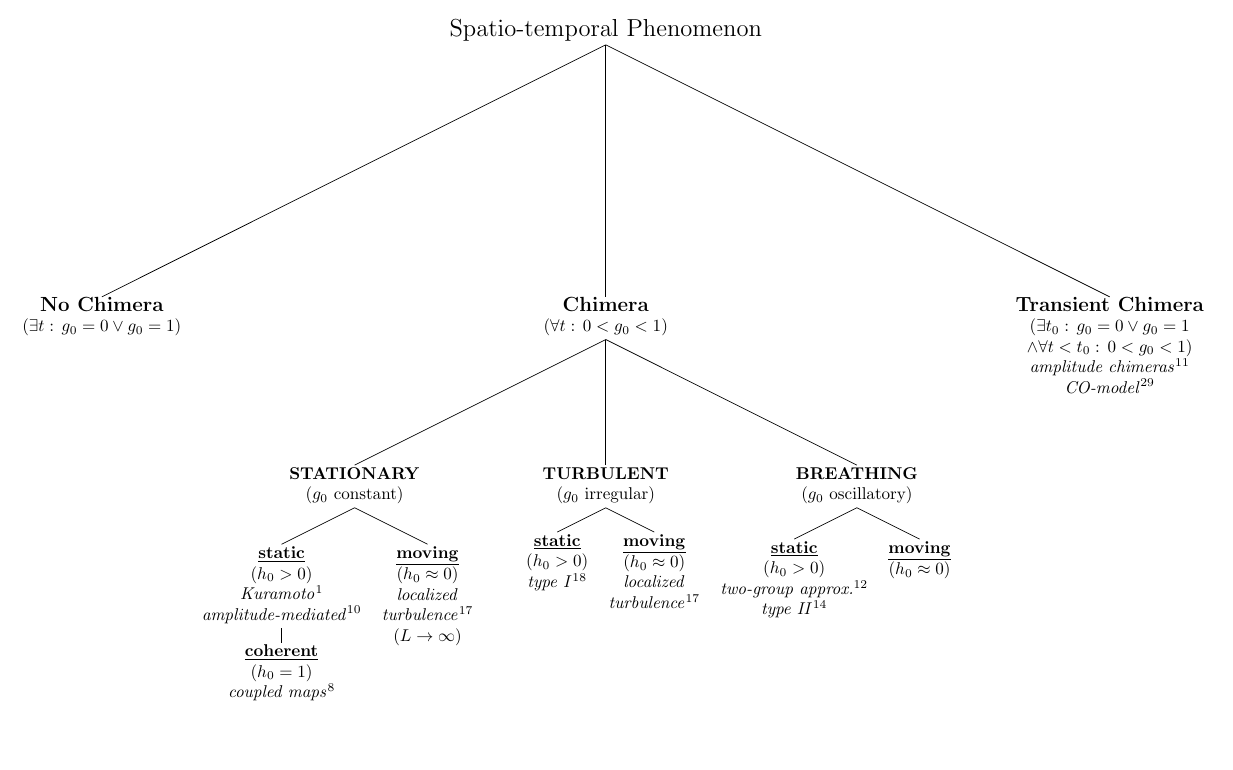}
% \resizebox{0.9\textwidth}{!}{%
% \Tree [.\Large{Spatio-temporal Phenomenon} [.{\large{\textbf{No Chimera}} \\
%   $(\exists t: \, g_0 = 0 \vee g_0 = 1)$ } ] 
% [.{\large{\textbf{Chimera}} \\ $(\forall t: \, 0<g_0<1)$}
% [.{\textbf{STATIONARY} \\ ($g_0 $ constant)}
% [.{\underline{\textbf{static}} \\ ($h_0 > 0$) \\ \textit{Kuramoto}{\protect{\cite{kuramoto2002}}} \\
%   \textit{amplitude-mediated}\cite{sethia2013}}
% {\underline{\textbf{coherent}} \\ ($h_0 = 1$) \\
%   \textit{coupled maps}\cite{hagerstrom2012} } ]
% {\underline{\textbf{moving}} \\ ($h_0 \approx 0$) \\ \textit{localized} \\ \textit{turbulence}\cite{Battogtokh1997} \\ $(L \rightarrow
%   \infty)$ } ]
% [.{\textbf{TURBULENT} \\ ($g_0$ irregular) } {
%   \underline{\textbf{static}} \\ ($h_0 > 0$) \\ \textit{type I}\cite{schmidt2015} } { \underline{\textbf{moving}} \\ ($h_0 \approx 0$) \\ \textit{localized} \\
%   \textit{turbulence}\cite{Battogtokh1997} } ] [.{\textbf{BREATHING}
%   \\ ($g_0 $ oscillatory)} {\underline{\textbf{static}} \\ ($h_0
%   > 0$) \\
%   \textit{two-group approx.}\cite{abrams2008} \\ \textit{type
%     II}\cite{schmidt2014} } ] ] 
%  [.{\large{\textbf{Transient Chimera}} \\
%   $(\exists t_0: \, g_0 = 0 \vee g_0 = 1$ \\ $\wedge \forall t<t_0: \,
%   0<g_0<1)$ \\ \textit{amplitude chimeras}\cite{zakharova2014} \\ \textit{CO-model}\cite{Falcke1995}} ] ] }
\end{figure*}
In figure \ref{fig:13}b, the behavior of $g_0(t)$ with time and the value of $h_0$
are shown. They are remarkably similar to the type II dynamics as
depicted in figure \ref{fig:7}. 
Hence we can conclude that the
observed experimental chimera is of the \textit{breathing} type.
The smallness of $h_0$ originates from
the fact that the coherent cluster is relatively small.

\section{Classification Scheme}
\label{sec:classification}

Above, we introduced two correlation measures, $g_0(t)$ and $h_0$,
which allow a quantification of coherence and incoherence in dynamical
systems. For phase oscillators, the local Kuramoto order parameter
already quantifies the degree of incoherence as a function of space
and time. In contrast, our global measure $g_0(t)$ yields information about the total relative sizes of the coherent and
incoherent parts of the system, but does not contain information about
local properties within the incoherent group.
Nevertheless, it exhibits
% We demonstrated above that the correlation measure $g_0(t)$ exhibits
distinct qualitative types of temporal 
behavior for chimera states with
visibly different dynamic features,
and thus, like the local order parameter, can be used to discriminate between chimeras, transient
chimeras and other types of dynamics. Its main advantage is its unrestricted
applicability, not only to ensembles of phase oscillators, but to any type of dynamical system.
Thus, $g_0(t)$
allows for a simple and
straightforward classification of general chimera states.\par
For $g_0(t)$ equal to $0$ or $1$, one of the two
phases, the coherent ($g_0(t)=0$) or incoherent one ($g_0(t)=1$), does
not exist.
This contradicts the requirement of 'coexistence', and we argue
that dynamical states where this occurs should be differentiated from
chimera states. This includes spatio-temporal intermittency, the
turbulent patterns in the CO model and the amplitude-chimeras shown in figure \ref{fig:10}.
Yet, for the latter two, $0<g_0(t)<1$ is valid for
a long time interval. Therefore, we suggest that these states are
categorized as \textit{transient} chimeras. In the case of intermittency,
$g_0(t)$ fluctuates constantly, thereby attaining a value of 0 after
arbitrary periods of time. It is therefore differentiated from chimera
states.\par
Chimera states, i.e. 
states with $0<g_0(t)<1$, can then be classified into three
groups:
\begin{enumerate}
\item \textbf{Stationary chimeras}: Chimera states with constant coherent cluster size
  $g_0(t)$,
\item \textbf{Turbulent chimeras}: Chimera states where the temporal evolution of $g_0(t)$
  is irregular,
\item \textbf{Breathing chimeras}: States in which the behavior of $g_0(t)$ is periodic.
\end{enumerate}
Note that there might be some ambiguity in the assignment to these
sub-categories, since the boundaries
between stationary/turbulent and turbulent/oscillatory are
rather fluent.\par
Based on the temporal correlation measure $h_0$, these groups can  be further divided into
three subclasses:
\begin{enumerate}[label=(\alph*)]
\item \textit{Static chimeras}, in which the coherent cluster is
  confined to the same position in space over time. That means, $h_0$ is non-zero and independent of the time window evaluated.
\item \textit{Moving chimeras}, where $h_0$ vanishes if the
  regarded time window is taken sufficiently large.
\item \textit{Time-coherent chimeras}, that is chimera states with no temporal
  incoherence and thus $h_0=1$.
\end{enumerate}
These criteria are summarized in a chimera classification scheme shown
in figure \ref{fig:tree}. The examples discussed in the last two
sections are assigned accordingly in the classification tree.\par
In conclusion, we have introduced two observables, $g_0(t)$ and $h_0$, that are
a measure for the degree of spatial and temporal coherence,
respectively, and allow for a discrimination between different types
of chimeras from simulated or experimental spatio-temporal data sets.
All examples from literature considered here could be assigned
to one of the classes. 
We verified the generality of the approach with additional examples,
such as the FitzHugh-Nagumo\cite{omel13} and R\"{o}ssler models\cite{omel12ro}.
Note, however, the scheme does not distinguish
between single- and multi-headed chimeras. Furthermore, it is likely that future studies will
reveal additional phenomena which the method does not account for at
the current stage. However, even in this
case, the classification scheme should present a useful base skeleton
that can be expanded as new discoveries will dictate. 

\begin{acknowledgments}
We thank Maximilian Patzauer and Konrad Sch\"onleber for providing
the experimental data and for fruitful discussions.
Financial support from the 
\textit{Institute for Advanced Study - Technische Universit\"{a}t
M\"{u}nchen},
funded by the German Excellence Initiative,
and the cluster of excellence \textit{Nanosystems Initiative Munich (NIM)}
is gratefully acknowledged.
\end{acknowledgments}

% \bibliography{lit}

%

%\bibliography{lit}

\begin{thebibliography}{34}%
\makeatletter
\providecommand \@ifxundefined [1]{%
 \@ifx{#1\undefined}
}%
\providecommand \@ifnum [1]{%
 \ifnum #1\expandafter \@firstoftwo
 \else \expandafter \@secondoftwo
 \fi
}%
\providecommand \@ifx [1]{%
 \ifx #1\expandafter \@firstoftwo
 \else \expandafter \@secondoftwo
 \fi
}%
\providecommand \natexlab [1]{#1}%
\providecommand \enquote  [1]{``#1''}%
\providecommand \bibnamefont  [1]{#1}%
\providecommand \bibfnamefont [1]{#1}%
\providecommand \citenamefont [1]{#1}%
\providecommand \href@noop [0]{\@secondoftwo}%
\providecommand \href [0]{\begingroup \@sanitize@url \@href}%
\providecommand \@href[1]{\@@startlink{#1}\@@href}%
\providecommand \@@href[1]{\endgroup#1\@@endlink}%
\providecommand \@sanitize@url [0]{\catcode `\\12\catcode `\$12\catcode
  `\&12\catcode `\#12\catcode `\^12\catcode `\_12\catcode `\%12\relax}%
\providecommand \@@startlink[1]{}%
\providecommand \@@endlink[0]{}%
\providecommand \url  [0]{\begingroup\@sanitize@url \@url }%
\providecommand \@url [1]{\endgroup\@href {#1}{\urlprefix }}%
\providecommand \urlprefix  [0]{URL }%
\providecommand \Eprint [0]{\href }%
\providecommand \doibase [0]{http://dx.doi.org/}%
\providecommand \selectlanguage [0]{\@gobble}%
\providecommand \bibinfo  [0]{\@secondoftwo}%
\providecommand \bibfield  [0]{\@secondoftwo}%
\providecommand \translation [1]{[#1]}%
\providecommand \BibitemOpen [0]{}%
\providecommand \bibitemStop [0]{}%
\providecommand \bibitemNoStop [0]{.\EOS\space}%
\providecommand \EOS [0]{\spacefactor3000\relax}%
\providecommand \BibitemShut  [1]{\csname bibitem#1\endcsname}%
\let\auto@bib@innerbib\@empty
%</preamble>
\bibitem [{\citenamefont {Kuramoto}\ and\ \citenamefont
  {Battogtokh}(2002)}]{kuramoto2002}%
  \BibitemOpen
  \bibfield  {author} {\bibinfo {author} {\bibfnamefont {Y.}~\bibnamefont
  {Kuramoto}}\ and\ \bibinfo {author} {\bibfnamefont {D.}~\bibnamefont
  {Battogtokh}},\ }\bibfield  {title} {\enquote {\bibinfo {title} {Coexistence
  of coherence and incoherence in nonlocally coupled phase oscillators},}\
  }\href@noop {} {\bibfield  {journal} {\bibinfo  {journal} {Nonlinear Phenom.
  Complex Syst.}\ }\textbf {\bibinfo {volume} {5}},\ \bibinfo {pages}
  {380--385} (\bibinfo {year} {2002})}\BibitemShut {NoStop}%
\bibitem [{\citenamefont {Abrams}\ and\ \citenamefont
  {Strogatz}(2004)}]{abrams2004}%
  \BibitemOpen
  \bibfield  {author} {\bibinfo {author} {\bibfnamefont {D.~M.}\ \bibnamefont
  {Abrams}}\ and\ \bibinfo {author} {\bibfnamefont {S.~H.}\ \bibnamefont
  {Strogatz}},\ }\bibfield  {title} {\enquote {\bibinfo {title} {Chimera states
  for coupled oscillators},}\ }\href@noop {} {\bibfield  {journal} {\bibinfo
  {journal} {Phys. Rev. Lett.}\ }\textbf {\bibinfo {volume} {93}},\ \bibinfo
  {pages} {174102} (\bibinfo {year} {2004})}\BibitemShut {NoStop}%
\bibitem [{\citenamefont {Laing}\ and\ \citenamefont {Chow}(2001)}]{laing}%
  \BibitemOpen
  \bibfield  {author} {\bibinfo {author} {\bibfnamefont {C.~R.}\ \bibnamefont
  {Laing}}\ and\ \bibinfo {author} {\bibfnamefont {C.~C.}\ \bibnamefont
  {Chow}},\ }\bibfield  {title} {\enquote {\bibinfo {title} {Stationary bumps
  in networks of spiking neurons},}\ }\href@noop {} {\bibfield  {journal}
  {\bibinfo  {journal} {Neural Computation}\ }\textbf {\bibinfo {volume}
  {12}},\ \bibinfo {pages} {1473--1494} (\bibinfo {year} {2001})}\BibitemShut
  {NoStop}%
\bibitem [{\citenamefont {Rattenborg}, \citenamefont {Amlaner},\ and\
  \citenamefont {Lima}(2000)}]{Rattenborg2000}%
  \BibitemOpen
  \bibfield  {author} {\bibinfo {author} {\bibfnamefont {N.}~\bibnamefont
  {Rattenborg}}, \bibinfo {author} {\bibfnamefont {C.}~\bibnamefont {Amlaner}},
  \ and\ \bibinfo {author} {\bibfnamefont {S.}~\bibnamefont {Lima}},\
  }\bibfield  {title} {\enquote {\bibinfo {title} {Behavioral,
  neurophysiological and evolutionary perspectives on unihemispheric sleep},}\
  }\href {\doibase http://dx.doi.org/10.1016/S0149-7634(00)00039-7} {\bibfield
  {journal} {\bibinfo  {journal} {Neuroscience and Biobehavioral Reviews}\
  }\textbf {\bibinfo {volume} {24}},\ \bibinfo {pages} {817 -- 842} (\bibinfo
  {year} {2000})}\BibitemShut {NoStop}%
\bibitem [{\citenamefont {Barkley}\ and\ \citenamefont
  {Tuckerman}(2005)}]{Barkley2005}%
  \BibitemOpen
  \bibfield  {author} {\bibinfo {author} {\bibfnamefont {D.}~\bibnamefont
  {Barkley}}\ and\ \bibinfo {author} {\bibfnamefont {L.~S.}\ \bibnamefont
  {Tuckerman}},\ }\bibfield  {title} {\enquote {\bibinfo {title} {Computational
  study of turbulent laminar patterns in couette flow},}\ }\href {\doibase
  10.1103/PhysRevLett.94.014502} {\bibfield  {journal} {\bibinfo  {journal}
  {Phys. Rev. Lett.}\ }\textbf {\bibinfo {volume} {94}},\ \bibinfo {pages}
  {014502} (\bibinfo {year} {2005})}\BibitemShut {NoStop}%
\bibitem [{\citenamefont {Duguet}\ and\ \citenamefont
  {Schlatter}(2013)}]{Duguet2013}%
  \BibitemOpen
  \bibfield  {author} {\bibinfo {author} {\bibfnamefont {Y.}~\bibnamefont
  {Duguet}}\ and\ \bibinfo {author} {\bibfnamefont {P.}~\bibnamefont
  {Schlatter}},\ }\bibfield  {title} {\enquote {\bibinfo {title} {Oblique
  laminar-turbulent interfaces in plane shear flows},}\ }\href {\doibase
  10.1103/PhysRevLett.110.034502} {\bibfield  {journal} {\bibinfo  {journal}
  {Phys. Rev. Lett.}\ }\textbf {\bibinfo {volume} {110}},\ \bibinfo {pages}
  {034502} (\bibinfo {year} {2013})}\BibitemShut {NoStop}%
\bibitem [{\citenamefont {Bordyugov}, \citenamefont {Pikovsky},\ and\
  \citenamefont {Rosenblum}(2010)}]{bordyugov2010}%
  \BibitemOpen
  \bibfield  {author} {\bibinfo {author} {\bibfnamefont {G.}~\bibnamefont
  {Bordyugov}}, \bibinfo {author} {\bibfnamefont {A.}~\bibnamefont {Pikovsky}},
  \ and\ \bibinfo {author} {\bibfnamefont {M.}~\bibnamefont {Rosenblum}},\
  }\bibfield  {title} {\enquote {\bibinfo {title} {Self-emerging and turbulent
  chimeras in oscillator chains},}\ }\href@noop {} {\bibfield  {journal}
  {\bibinfo  {journal} {Phys. Rev. E}\ }\textbf {\bibinfo {volume} {82}},\
  \bibinfo {pages} {035205} (\bibinfo {year} {2010})}\BibitemShut {NoStop}%
\bibitem [{\citenamefont {Omelchenko}\ \emph {et~al.}(2011)\citenamefont
  {Omelchenko}, \citenamefont {Maistrenko}, \citenamefont {H{\"{o}}vel},\ and\
  \citenamefont {Sch{\"{o}}ll}}]{omelchenko2011}%
  \BibitemOpen
  \bibfield  {author} {\bibinfo {author} {\bibfnamefont {I.}~\bibnamefont
  {Omelchenko}}, \bibinfo {author} {\bibfnamefont {Y.}~\bibnamefont
  {Maistrenko}}, \bibinfo {author} {\bibfnamefont {P.}~\bibnamefont
  {H{\"{o}}vel}}, \ and\ \bibinfo {author} {\bibfnamefont {E.}~\bibnamefont
  {Sch{\"{o}}ll}},\ }\bibfield  {title} {\enquote {\bibinfo {title} {{Loss of
  Coherence in Dynamical Networks: Spatial Chaos and Chimera States}},}\ }\href
  {\doibase 10.1103/PhysRevLett.106.234102} {\bibfield  {journal} {\bibinfo
  {journal} {Physical Review Letters}\ }\textbf {\bibinfo {volume} {106}},\
  \bibinfo {pages} {234102} (\bibinfo {year} {2011})}\BibitemShut {NoStop}%
\bibitem [{\citenamefont {Hagerstrom}\ \emph {et~al.}(2012)\citenamefont
  {Hagerstrom}, \citenamefont {Murphy}, \citenamefont {Roy}, \citenamefont
  {H\"{o}vel}, \citenamefont {Omelchenko},\ and\ \citenamefont
  {Sch\"{o}ll}}]{hagerstrom2012}%
  \BibitemOpen
  \bibfield  {author} {\bibinfo {author} {\bibfnamefont {A.~M.}\ \bibnamefont
  {Hagerstrom}}, \bibinfo {author} {\bibfnamefont {T.~E.}\ \bibnamefont
  {Murphy}}, \bibinfo {author} {\bibfnamefont {R.}~\bibnamefont {Roy}},
  \bibinfo {author} {\bibfnamefont {P.}~\bibnamefont {H\"{o}vel}}, \bibinfo
  {author} {\bibfnamefont {I.}~\bibnamefont {Omelchenko}}, \ and\ \bibinfo
  {author} {\bibfnamefont {E.}~\bibnamefont {Sch\"{o}ll}},\ }\bibfield  {title}
  {\enquote {\bibinfo {title} {Experimental observation of chimeras in
  coupled-map lattices},}\ }\href@noop {} {\bibfield  {journal} {\bibinfo
  {journal} {Nature Physics}\ }\textbf {\bibinfo {volume} {8}},\ \bibinfo
  {pages} {658--661} (\bibinfo {year} {2012})}\BibitemShut {NoStop}%
\bibitem [{\citenamefont {Omelchenko}\ \emph
  {et~al.}(2013{\natexlab{a}})\citenamefont {Omelchenko}, \citenamefont
  {Omel'chenko}, \citenamefont {H\"ovel},\ and\ \citenamefont
  {Sch\"oll}}]{omelchenk2013}%
  \BibitemOpen
  \bibfield  {author} {\bibinfo {author} {\bibfnamefont {I.}~\bibnamefont
  {Omelchenko}}, \bibinfo {author} {\bibfnamefont {O.~E.}\ \bibnamefont
  {Omel'chenko}}, \bibinfo {author} {\bibfnamefont {P.}~\bibnamefont
  {H\"ovel}}, \ and\ \bibinfo {author} {\bibfnamefont {E.}~\bibnamefont
  {Sch\"oll}},\ }\bibfield  {title} {\enquote {\bibinfo {title} {When nonlocal
  coupling between oscillators becomes stronger: patched synchrony or
  multichimera states},}\ }\href@noop {} {\bibfield  {journal} {\bibinfo
  {journal} {Phys. Rev. Lett.}\ }\textbf {\bibinfo {volume} {110}},\ \bibinfo
  {pages} {224101} (\bibinfo {year} {2013}{\natexlab{a}})}\BibitemShut
  {NoStop}%
\bibitem [{\citenamefont {Sethia}, \citenamefont {Sen},\ and\ \citenamefont
  {Johnston}(2013)}]{sethia2013}%
  \BibitemOpen
  \bibfield  {author} {\bibinfo {author} {\bibfnamefont {G.~C.}\ \bibnamefont
  {Sethia}}, \bibinfo {author} {\bibfnamefont {A.}~\bibnamefont {Sen}}, \ and\
  \bibinfo {author} {\bibfnamefont {G.~L.}\ \bibnamefont {Johnston}},\
  }\bibfield  {title} {\enquote {\bibinfo {title} {Amplitude-mediated chimera
  states},}\ }\href@noop {} {\bibfield  {journal} {\bibinfo  {journal} {Phys.
  Rev. E}\ }\textbf {\bibinfo {volume} {88}},\ \bibinfo {pages} {042917}
  (\bibinfo {year} {2013})}\BibitemShut {NoStop}%
\bibitem [{\citenamefont {Zakharova}, \citenamefont {Kapeller},\ and\
  \citenamefont {Sch\"oll}(2014)}]{zakharova2014}%
  \BibitemOpen
  \bibfield  {author} {\bibinfo {author} {\bibfnamefont {A.}~\bibnamefont
  {Zakharova}}, \bibinfo {author} {\bibfnamefont {M.}~\bibnamefont {Kapeller}},
  \ and\ \bibinfo {author} {\bibfnamefont {E.}~\bibnamefont {Sch\"oll}},\
  }\bibfield  {title} {\enquote {\bibinfo {title} {Chimera death: symmetry
  breaking in dynamical networks},}\ }\href@noop {} {\bibfield  {journal}
  {\bibinfo  {journal} {Phys. Rev. Lett.}\ }\textbf {\bibinfo {volume} {112}},\
  \bibinfo {pages} {154101} (\bibinfo {year} {2014})}\BibitemShut {NoStop}%
\bibitem [{\citenamefont {Abrams}\ \emph {et~al.}(2008)\citenamefont {Abrams},
  \citenamefont {Mirollo}, \citenamefont {Strogatz},\ and\ \citenamefont
  {Wiley}}]{abrams2008}%
  \BibitemOpen
  \bibfield  {author} {\bibinfo {author} {\bibfnamefont {D.~M.}\ \bibnamefont
  {Abrams}}, \bibinfo {author} {\bibfnamefont {R.}~\bibnamefont {Mirollo}},
  \bibinfo {author} {\bibfnamefont {S.~H.}\ \bibnamefont {Strogatz}}, \ and\
  \bibinfo {author} {\bibfnamefont {D.~A.}\ \bibnamefont {Wiley}},\ }\bibfield
  {title} {\enquote {\bibinfo {title} {Solvable model for chimera states of
  coupled oscillators},}\ }\href@noop {} {\bibfield  {journal} {\bibinfo
  {journal} {Phys. Rev. Lett.}\ }\textbf {\bibinfo {volume} {101}},\ \bibinfo
  {pages} {084103} (\bibinfo {year} {2008})}\BibitemShut {NoStop}%
\bibitem [{\citenamefont {Tinsley}, \citenamefont {Nkomo},\ and\ \citenamefont
  {Showalter}(2012)}]{tinsley2012}%
  \BibitemOpen
  \bibfield  {author} {\bibinfo {author} {\bibfnamefont {M.~R.}\ \bibnamefont
  {Tinsley}}, \bibinfo {author} {\bibfnamefont {S.}~\bibnamefont {Nkomo}}, \
  and\ \bibinfo {author} {\bibfnamefont {K.}~\bibnamefont {Showalter}},\
  }\bibfield  {title} {\enquote {\bibinfo {title} {Chimera and phase-cluster
  states in populations of coupled chemical oscillators},}\ }\href@noop {}
  {\bibfield  {journal} {\bibinfo  {journal} {Nature Physics}\ }\textbf
  {\bibinfo {volume} {8}},\ \bibinfo {pages} {662--665} (\bibinfo {year}
  {2012})}\BibitemShut {NoStop}%
\bibitem [{\citenamefont {Schmidt}\ \emph {et~al.}(2014)\citenamefont
  {Schmidt}, \citenamefont {Sch\"onleber}, \citenamefont {Krischer},\ and\
  \citenamefont {Garc\'ia-Morales}}]{schmidt2014}%
  \BibitemOpen
  \bibfield  {author} {\bibinfo {author} {\bibfnamefont {L.}~\bibnamefont
  {Schmidt}}, \bibinfo {author} {\bibfnamefont {K.}~\bibnamefont
  {Sch\"onleber}}, \bibinfo {author} {\bibfnamefont {K.}~\bibnamefont
  {Krischer}}, \ and\ \bibinfo {author} {\bibfnamefont {V.}~\bibnamefont
  {Garc\'ia-Morales}},\ }\bibfield  {title} {\enquote {\bibinfo {title}
  {Coexistence of synchrony and incoherence in oscillatory media under
  nonlinear global coupling},}\ }\href@noop {} {\bibfield  {journal} {\bibinfo
  {journal} {Chaos}\ }\textbf {\bibinfo {volume} {24}},\ \bibinfo {eid}
  {013102} (\bibinfo {year} {2014})}\BibitemShut {NoStop}%
\bibitem [{\citenamefont {Yeldesbay}, \citenamefont {Pikovsky},\ and\
  \citenamefont {Rosenblum}(2014)}]{yeldesbay2014}%
  \BibitemOpen
  \bibfield  {author} {\bibinfo {author} {\bibfnamefont {A.}~\bibnamefont
  {Yeldesbay}}, \bibinfo {author} {\bibfnamefont {A.}~\bibnamefont {Pikovsky}},
  \ and\ \bibinfo {author} {\bibfnamefont {M.}~\bibnamefont {Rosenblum}},\
  }\bibfield  {title} {\enquote {\bibinfo {title} {Chimeralike states in an
  ensemble of globally coupled oscillators},}\ }\href@noop {} {\bibfield
  {journal} {\bibinfo  {journal} {Phys. Rev. Lett.}\ }\textbf {\bibinfo
  {volume} {112}} (\bibinfo {year} {2014})}\BibitemShut {NoStop}%
\bibitem [{\citenamefont {Martens}\ \emph {et~al.}(2013)\citenamefont
  {Martens}, \citenamefont {Thutupalli}, \citenamefont {Fourrière},\ and\
  \citenamefont {Hallatschek}}]{martens}%
  \BibitemOpen
  \bibfield  {author} {\bibinfo {author} {\bibfnamefont {E.~A.}\ \bibnamefont
  {Martens}}, \bibinfo {author} {\bibfnamefont {S.}~\bibnamefont {Thutupalli}},
  \bibinfo {author} {\bibfnamefont {A.}~\bibnamefont {Fourrière}}, \ and\
  \bibinfo {author} {\bibfnamefont {O.}~\bibnamefont {Hallatschek}},\
  }\bibfield  {title} {\enquote {\bibinfo {title} {Chimera states in mechanical
  oscillator networks},}\ }\href {\doibase 10.1073/pnas.1302880110} {\bibfield
  {journal} {\bibinfo  {journal} {Proceedings of the National Academy of
  Sciences}\ }\textbf {\bibinfo {volume} {110}},\ \bibinfo {pages}
  {10563--10567} (\bibinfo {year} {2013})}\BibitemShut {NoStop}%
\bibitem [{\citenamefont {Battogtokh}, \citenamefont {Preusser},\ and\
  \citenamefont {Mikhailov}(1997)}]{Battogtokh1997}%
  \BibitemOpen
  \bibfield  {author} {\bibinfo {author} {\bibfnamefont {D.}~\bibnamefont
  {Battogtokh}}, \bibinfo {author} {\bibfnamefont {A.}~\bibnamefont
  {Preusser}}, \ and\ \bibinfo {author} {\bibfnamefont {A.}~\bibnamefont
  {Mikhailov}},\ }\bibfield  {title} {\enquote {\bibinfo {title} {Controlling
  turbulence in the complex {G}inzburg-{L}andau equation {II}.
  {T}wo-dimensional systems},}\ }\href@noop {} {\bibfield  {journal} {\bibinfo
  {journal} {Physica D: Nonlinear Phenomena}\ }\textbf {\bibinfo {volume}
  {106}},\ \bibinfo {pages} {327--362} (\bibinfo {year} {1997})}\BibitemShut
  {NoStop}%
\bibitem [{\citenamefont {Schmidt}\ and\ \citenamefont
  {Krischer}(2015{\natexlab{a}})}]{schmidt2015}%
  \BibitemOpen
  \bibfield  {author} {\bibinfo {author} {\bibfnamefont {L.}~\bibnamefont
  {Schmidt}}\ and\ \bibinfo {author} {\bibfnamefont {K.}~\bibnamefont
  {Krischer}},\ }\bibfield  {title} {\enquote {\bibinfo {title} {Chimeras in
  globally coupled oscillatory systems: from ensembles of oscillators to
  spatially continuous media},}\ }\href@noop {} {\bibfield  {journal} {\bibinfo
   {journal} {Chaos}\ }\textbf {\bibinfo {volume} {25}},\ \bibinfo {eid}
  {064401} (\bibinfo {year} {2015}{\natexlab{a}})}\BibitemShut {NoStop}%
\bibitem [{\citenamefont {Wolfrum}\ and\ \citenamefont
  {Omel'chenko}(2011)}]{wolfrum2011}%
  \BibitemOpen
  \bibfield  {author} {\bibinfo {author} {\bibfnamefont {M.}~\bibnamefont
  {Wolfrum}}\ and\ \bibinfo {author} {\bibfnamefont {O.~E.}\ \bibnamefont
  {Omel'chenko}},\ }\bibfield  {title} {\enquote {\bibinfo {title} {Chimera
  states are chaotic transients},}\ }\href@noop {} {\bibfield  {journal}
  {\bibinfo  {journal} {Phys. Rev. E}\ }\textbf {\bibinfo {volume} {84}},\
  \bibinfo {pages} {015201} (\bibinfo {year} {2011})}\BibitemShut {NoStop}%
\bibitem [{\citenamefont {Ashwin}\ and\ \citenamefont
  {Burylko}(2015)}]{ashwin2015}%
  \BibitemOpen
  \bibfield  {author} {\bibinfo {author} {\bibfnamefont {P.}~\bibnamefont
  {Ashwin}}\ and\ \bibinfo {author} {\bibfnamefont {O.}~\bibnamefont
  {Burylko}},\ }\bibfield  {title} {\enquote {\bibinfo {title} {Weak chimeras
  in minimal networks of coupled phase oscillators},}\ }\href@noop {}
  {\bibfield  {journal} {\bibinfo  {journal} {Chaos}\ }\textbf {\bibinfo
  {volume} {25}},\ \bibinfo {eid} {013106} (\bibinfo {year}
  {2015})}\BibitemShut {NoStop}%
\bibitem [{\citenamefont {Schmidt}\ and\ \citenamefont
  {Krischer}(2015{\natexlab{b}})}]{Schmidt2015_2}%
  \BibitemOpen
  \bibfield  {author} {\bibinfo {author} {\bibfnamefont {L.}~\bibnamefont
  {Schmidt}}\ and\ \bibinfo {author} {\bibfnamefont {K.}~\bibnamefont
  {Krischer}},\ }\bibfield  {title} {\enquote {\bibinfo {title} {Clustering as
  a prerequisite for chimera states in globally coupled systems},}\ }\href
  {\doibase 10.1103/PhysRevLett.114.034101} {\bibfield  {journal} {\bibinfo
  {journal} {Phys. Rev. Lett.}\ }\textbf {\bibinfo {volume} {114}},\ \bibinfo
  {pages} {034101} (\bibinfo {year} {2015}{\natexlab{b}})}\BibitemShut
  {NoStop}%
\bibitem [{\citenamefont {Panaggio}\ and\ \citenamefont
  {Abrams}(2015)}]{panaggio2015}%
  \BibitemOpen
  \bibfield  {author} {\bibinfo {author} {\bibfnamefont {M.~J.}\ \bibnamefont
  {Panaggio}}\ and\ \bibinfo {author} {\bibfnamefont {D.~M.}\ \bibnamefont
  {Abrams}},\ }\bibfield  {title} {\enquote {\bibinfo {title} {Chimera states:
  coexistence of coherence and incoherence in networks of coupled
  oscillators},}\ }\href@noop {} {\bibfield  {journal} {\bibinfo  {journal}
  {Nonlinearity}\ }\textbf {\bibinfo {volume} {28}},\ \bibinfo {pages}
  {R67--R87} (\bibinfo {year} {2015})}\BibitemShut {NoStop}%
\bibitem [{\citenamefont {Gopal}\ \emph {et~al.}(2014)\citenamefont {Gopal},
  \citenamefont {Chandrasekar}, \citenamefont {Venkatesan},\ and\ \citenamefont
  {Lakshmanan}}]{gopal2014}%
  \BibitemOpen
  \bibfield  {author} {\bibinfo {author} {\bibfnamefont {R.}~\bibnamefont
  {Gopal}}, \bibinfo {author} {\bibfnamefont {V.~K.}\ \bibnamefont
  {Chandrasekar}}, \bibinfo {author} {\bibfnamefont {A.}~\bibnamefont
  {Venkatesan}}, \ and\ \bibinfo {author} {\bibfnamefont {M.}~\bibnamefont
  {Lakshmanan}},\ }\bibfield  {title} {\enquote {\bibinfo {title} {Observation
  and characterization of chimera states in coupled dynamical systems with
  nonlocal coupling},}\ }\href@noop {} {\bibfield  {journal} {\bibinfo
  {journal} {Phys. Rev. E}\ }\textbf {\bibinfo {volume} {89}},\ \bibinfo
  {pages} {052914} (\bibinfo {year} {2014})}\BibitemShut {NoStop}%
\bibitem [{sup()}]{supp}%
  \BibitemOpen
  \href@noop {} {}\bibinfo {note} {See supplemental material at [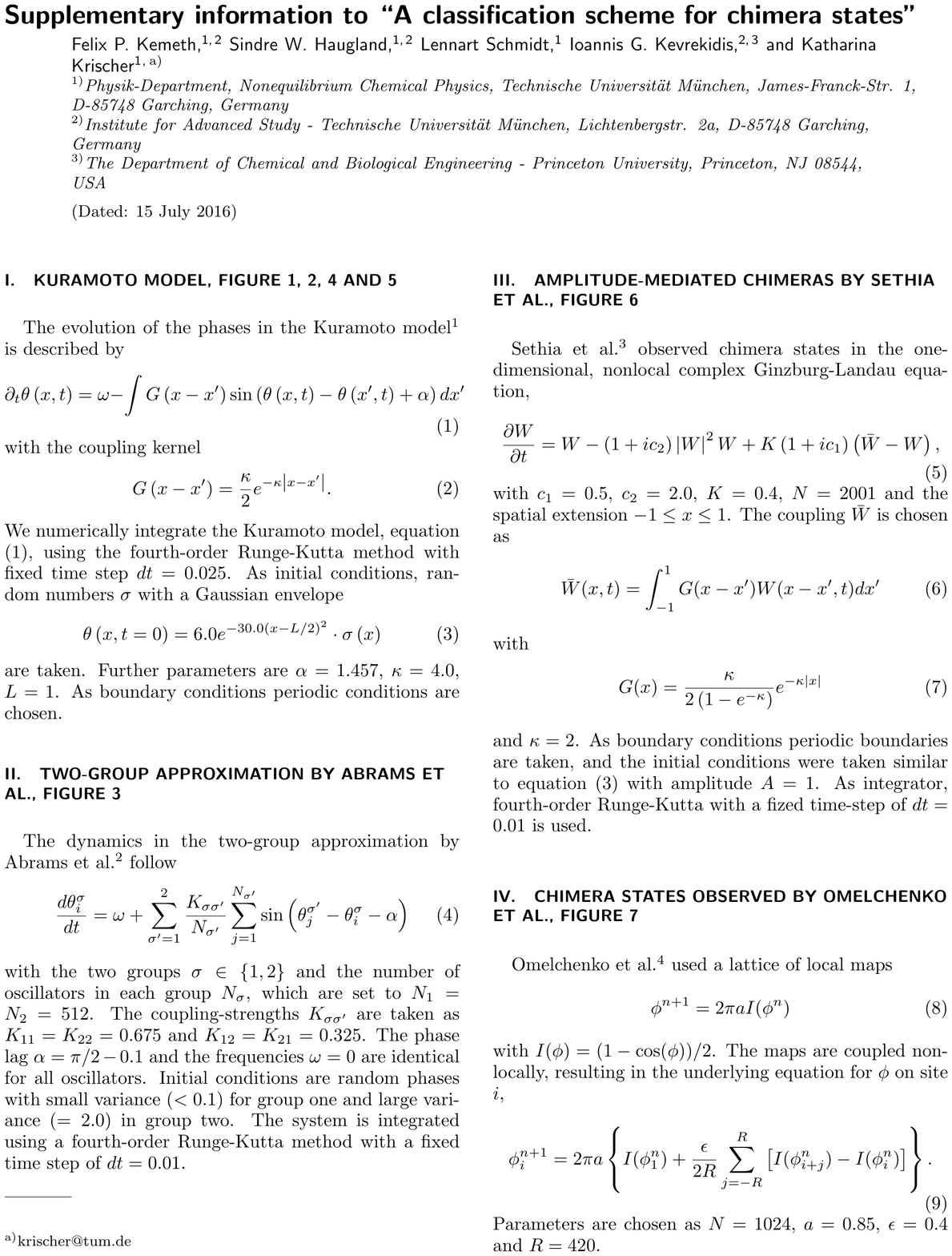] for
  details on to the individual systems and on the numerical methods
  used}\BibitemShut {NoStop}%
\bibitem [{\citenamefont {Shraiman}\ \emph {et~al.}(1992)\citenamefont
  {Shraiman}, \citenamefont {Pumir}, \citenamefont {van Saarloos},
  \citenamefont {Hohenberg}, \citenamefont {Chaté},\ and\ \citenamefont
  {Holen}}]{shraiman1992}%
  \BibitemOpen
  \bibfield  {author} {\bibinfo {author} {\bibfnamefont {B.}~\bibnamefont
  {Shraiman}}, \bibinfo {author} {\bibfnamefont {A.}~\bibnamefont {Pumir}},
  \bibinfo {author} {\bibfnamefont {W.}~\bibnamefont {van Saarloos}}, \bibinfo
  {author} {\bibfnamefont {P.}~\bibnamefont {Hohenberg}}, \bibinfo {author}
  {\bibfnamefont {H.}~\bibnamefont {Chaté}}, \ and\ \bibinfo {author}
  {\bibfnamefont {M.}~\bibnamefont {Holen}},\ }\bibfield  {title} {\enquote
  {\bibinfo {title} {Spatiotemporal chaos in the one-dimensional complex
  {G}inzburg-{L}andau equation},}\ }\href@noop {} {\bibfield  {journal}
  {\bibinfo  {journal} {Physica D: Nonlinear Phenomena}\ }\textbf {\bibinfo
  {volume} {57}},\ \bibinfo {pages} {241 -- 248} (\bibinfo {year}
  {1992})}\BibitemShut {NoStop}%
\bibitem [{\citenamefont {Haugland}, \citenamefont {Schmidt},\ and\
  \citenamefont {Krischer}(2015)}]{haugland2015}%
  \BibitemOpen
  \bibfield  {author} {\bibinfo {author} {\bibfnamefont {S.~W.}\ \bibnamefont
  {Haugland}}, \bibinfo {author} {\bibfnamefont {L.}~\bibnamefont {Schmidt}}, \
  and\ \bibinfo {author} {\bibfnamefont {K.}~\bibnamefont {Krischer}},\
  }\bibfield  {title} {\enquote {\bibinfo {title} {Self-organized alternating
  chimera states in oscillatory media},}\ }\href@noop {} {\bibfield  {journal}
  {\bibinfo  {journal} {Scientific Reports}\ }\textbf {\bibinfo {volume} {5}},\
  \bibinfo {pages} {9883} (\bibinfo {year} {2015})}\BibitemShut {NoStop}%
\bibitem [{\citenamefont {Loos}\ \emph {et~al.}(2016)\citenamefont {Loos},
  \citenamefont {Claussen}, \citenamefont {Sch\"oll},\ and\ \citenamefont
  {Zakharova}}]{zakharova2016}%
  \BibitemOpen
  \bibfield  {author} {\bibinfo {author} {\bibfnamefont {S.~A.~M.}\
  \bibnamefont {Loos}}, \bibinfo {author} {\bibfnamefont {J.~C.}\ \bibnamefont
  {Claussen}}, \bibinfo {author} {\bibfnamefont {E.}~\bibnamefont {Sch\"oll}},
  \ and\ \bibinfo {author} {\bibfnamefont {A.}~\bibnamefont {Zakharova}},\
  }\bibfield  {title} {\enquote {\bibinfo {title} {Chimera patterns under the
  impact of noise},}\ }\href@noop {} {\bibfield  {journal} {\bibinfo  {journal}
  {Phys. Rev. E}\ }\textbf {\bibinfo {volume} {93}},\ \bibinfo {pages} {012209}
  (\bibinfo {year} {2016})}\BibitemShut {NoStop}%
\bibitem [{\citenamefont {Falcke}\ and\ \citenamefont
  {Engel}(1994{\natexlab{a}})}]{Falcke1993}%
  \BibitemOpen
  \bibfield  {author} {\bibinfo {author} {\bibfnamefont {M.}~\bibnamefont
  {Falcke}}\ and\ \bibinfo {author} {\bibfnamefont {H.}~\bibnamefont {Engel}},\
  }\bibfield  {title} {\enquote {\bibinfo {title} {Influence of global coupling
  through the gas phase on the dynamics of {CO} oxidation on {P}t(110)},}\
  }\href {\doibase 10.1103/PhysRevE.50.1353} {\bibfield  {journal} {\bibinfo
  {journal} {Phys. Rev. E}\ }\textbf {\bibinfo {volume} {50}},\ \bibinfo
  {pages} {1353--1359} (\bibinfo {year} {1994}{\natexlab{a}})}\BibitemShut
  {NoStop}%
\bibitem [{\citenamefont {Falcke}\ and\ \citenamefont
  {Engel}(1994{\natexlab{b}})}]{Falcke1994}%
  \BibitemOpen
  \bibfield  {author} {\bibinfo {author} {\bibfnamefont {M.}~\bibnamefont
  {Falcke}}\ and\ \bibinfo {author} {\bibfnamefont {H.}~\bibnamefont {Engel}},\
  }\bibfield  {title} {\enquote {\bibinfo {title} {Pattern formation during the
  {CO} oxidation on {P}t(110) surfaces under global coupling},}\ }\href
  {\doibase http://dx.doi.org/10.1063/1.468379} {\bibfield  {journal} {\bibinfo
   {journal} {The Journal of Chemical Physics}\ }\textbf {\bibinfo {volume}
  {101}},\ \bibinfo {pages} {6255--6263} (\bibinfo {year}
  {1994}{\natexlab{b}})}\BibitemShut {NoStop}%
\bibitem [{\citenamefont {Falcke}(1995)}]{Falcke1995}%
  \BibitemOpen
  \bibfield  {author} {\bibinfo {author} {\bibfnamefont {M.}~\bibnamefont
  {Falcke}},\ }\href@noop {} {\emph {\bibinfo {title} {Strukturbildung in
  Reaktions- Diffusionssystemen und globale Kopplung}}}\ (\bibinfo  {publisher}
  {Wissenschaft und Technik Verlag Gross},\ \bibinfo {address} {Berlin,
  Sebastianstr. 84},\ \bibinfo {year} {1995})\BibitemShut {NoStop}%
\bibitem [{\citenamefont {Sch\"onleber}\ \emph {et~al.}(2014)\citenamefont
  {Sch\"onleber}, \citenamefont {Zensen}, \citenamefont {Heinrich},\ and\
  \citenamefont {Krischer}}]{Konrad2014}%
  \BibitemOpen
  \bibfield  {author} {\bibinfo {author} {\bibfnamefont {K.}~\bibnamefont
  {Sch\"onleber}}, \bibinfo {author} {\bibfnamefont {C.}~\bibnamefont
  {Zensen}}, \bibinfo {author} {\bibfnamefont {A.}~\bibnamefont {Heinrich}}, \
  and\ \bibinfo {author} {\bibfnamefont {K.}~\bibnamefont {Krischer}},\
  }\bibfield  {title} {\enquote {\bibinfo {title} {Pattern formation during the
  oscillatory photoelectrodissolution of n-type silicon: turbulence, clusters
  and chimeras},}\ }\href {http://stacks.iop.org/1367-2630/16/i=6/a=063024}
  {\bibfield  {journal} {\bibinfo  {journal} {New Journal of Physics}\ }\textbf
  {\bibinfo {volume} {16}},\ \bibinfo {pages} {063024} (\bibinfo {year}
  {2014})}\BibitemShut {NoStop}%
\bibitem [{\citenamefont {Omelchenko}\ \emph
  {et~al.}(2013{\natexlab{b}})\citenamefont {Omelchenko}, \citenamefont
  {Omel'chenko}, \citenamefont {H\"ovel},\ and\ \citenamefont
  {Sch\"oll}}]{omel13}%
  \BibitemOpen
  \bibfield  {author} {\bibinfo {author} {\bibfnamefont {I.}~\bibnamefont
  {Omelchenko}}, \bibinfo {author} {\bibfnamefont {O.~E.}\ \bibnamefont
  {Omel'chenko}}, \bibinfo {author} {\bibfnamefont {P.}~\bibnamefont
  {H\"ovel}}, \ and\ \bibinfo {author} {\bibfnamefont {E.}~\bibnamefont
  {Sch\"oll}},\ }\bibfield  {title} {\enquote {\bibinfo {title} {When nonlocal
  coupling between oscillators becomes stronger: Patched synchrony or
  multichimera states},}\ }\href {\doibase 10.1103/PhysRevLett.110.224101}
  {\bibfield  {journal} {\bibinfo  {journal} {Phys. Rev. Lett.}\ }\textbf
  {\bibinfo {volume} {110}},\ \bibinfo {pages} {224101} (\bibinfo {year}
  {2013}{\natexlab{b}})}\BibitemShut {NoStop}%
\bibitem [{\citenamefont {Omelchenko}\ \emph {et~al.}(2012)\citenamefont
  {Omelchenko}, \citenamefont {Riemenschneider}, \citenamefont {H\"ovel},
  \citenamefont {Maistrenko},\ and\ \citenamefont {Sch\"oll}}]{omel12ro}%
  \BibitemOpen
  \bibfield  {author} {\bibinfo {author} {\bibfnamefont {I.}~\bibnamefont
  {Omelchenko}}, \bibinfo {author} {\bibfnamefont {B.}~\bibnamefont
  {Riemenschneider}}, \bibinfo {author} {\bibfnamefont {P.}~\bibnamefont
  {H\"ovel}}, \bibinfo {author} {\bibfnamefont {Y.}~\bibnamefont {Maistrenko}},
  \ and\ \bibinfo {author} {\bibfnamefont {E.}~\bibnamefont {Sch\"oll}},\
  }\bibfield  {title} {\enquote {\bibinfo {title} {Transition from spatial
  coherence to incoherence in coupled chaotic systems},}\ }\href {\doibase
  10.1103/PhysRevE.85.026212} {\bibfield  {journal} {\bibinfo  {journal} {Phys.
  Rev. E}\ }\textbf {\bibinfo {volume} {85}},\ \bibinfo {pages} {026212}
  (\bibinfo {year} {2012})}\BibitemShut {NoStop}%
\end{thebibliography}%
\end{document}